\documentclass{article}

\usepackage{arxiv}
\usepackage[utf8]{inputenc}
\usepackage[T1]{fontenc}
\usepackage{hyperref}
\usepackage{url}
\usepackage{booktabs}
\usepackage{amsfonts}
\usepackage{nicefrac}
\usepackage{microtype}
\usepackage{graphicx}
\usepackage[numbers]{natbib}
\usepackage{doi}
\usepackage{float}
\usepackage{amsmath}
\usepackage{enumitem}

\def\x{\mathbf{x}}
\def\m{\mathbf{m}}
\def\f{\mathbf{f}}
\def\a{\mathbf{a}}
\def\v{\mathbf{v}}
\def\vp{\v_\text{tuned}}
\def\D{\mathcal{D}}
\def\S{\mathcal{S}}
\def\E{\mathcal{E}}
\def\C{\mathcal{C}}
\def\A{\mathcal{A}}
\def\P{\mathcal{P}}
\def\M{\mathcal{M}}
\def\R{\mathcal{R}}

\title{Spectrogram-Based Detection of Auto-Tuned Vocals in Music Recordings}

\date{March 8, 2024}

\author{
  \href{http://orcid.org/0000-0002-1021-709X}{\includegraphics[scale=0.06]{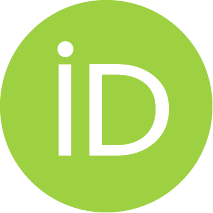}\hspace{1mm}Mahyar Gohari} \\
  Department of Information Engineering \\
  University of Brescia \\
  \And
  \href{http://orcid.org/0000-0003-0406-0222}{\includegraphics[scale=0.06]{orcid.pdf}\hspace{1mm}Paolo Bestagini} \\
  Department of Electronics, Information and Bioengineering \\
  Polytechnic University of Milan\\
  \And
  \href{http://orcid.org/0000-0003-2152-9424}{\includegraphics[scale=0.06]{orcid.pdf}\hspace{1mm}Sergio Benini} \\
  Department of Information Engineering \\
  University of Brescia \\
  \And
  \href{http://orcid.org/0000-0002-8879-9456}{\includegraphics[scale=0.06]{orcid.pdf}\hspace{1mm}Nicola Adami} \\
  Department of Information Engineering \\
  University of Brescia \\
}

\renewcommand{\shorttitle}{\textit{Spectrogram-Based Detection of Auto-Tuned Vocals in Music Recordings}}

\begin{document}
\maketitle

\begin{abstract}
In the domain of music production and audio processing, the implementation of automatic pitch correction of the singing voice, also known as Auto-Tune, has significantly transformed the landscape of vocal performance. While auto-tuning technology has offered musicians the ability to tune their vocal pitches and achieve a desired level of precision, its use has also sparked debates regarding its impact on authenticity and artistic integrity. As a result, detecting and analyzing Auto-Tuned vocals in music recordings has become essential for music scholars, producers, and listeners. However, to the best of our knowledge, no prior effort has been made in this direction. This study introduces a data-driven approach leveraging triplet networks for the detection of Auto-Tuned songs, backed by the creation of a dataset composed of original and Auto-Tuned audio clips. The experimental results demonstrate the superiority of the proposed method in both accuracy and robustness compared to Rawnet2, an end-to-end model proposed for anti-spoofing and widely used for other audio forensic tasks.
\end{abstract}


\section{Introduction}
Multimedia forgery detection refers to the process of identifying and detecting various types of manipulations or forgeries in multimedia content, such as images, audio, and video \cite{bhagtani2022}. Similarly, automated detection of forgeries in audio, and more specifically music, represents a crucial and active research area. This significance is increased by recent advancements in deep learning algorithms and the widespread availability of editing tools, highlighting the importance of exploration within this domain.
Forensic studies in Music Information Retrieval (MIR) have a long-standing history. Much research focuses on investigating the authenticity of music. In one of the earliest studies on music forgery detection by Li et al. \cite{wli2013}, mid-level chroma features and fuzzy logic were employed to detect various music manipulations such as cropping, adding, and replacing. However, recent research in this domain has shifted towards the more complex task of detecting plagiarized music \cite{deprisco2017, he2021, borkar2021, park2022, malandrino2022, lopezgarcia2022, gurjar2023}. For example, Gurjar et al. \cite{gurjar2023} introduced `TruMuzic', a model specifically designed to identify plagiarism in music. This is achieved through the utilization of provenance data and a deep learning-based similarity classification approach. 
However, these studies do not examine the authenticity of vocal pitch in music, which is the primary focus of this research. Auto-Tune, named after Antares Auto-Tune \cite{antares-autotune}, is a digital audio processing technique primarily employed to correct or manipulate pitch in music recordings. It functions by identifying and modifying the pitch of a vocalist's performance to achieve the desired musical output. The necessity of exploring this area arises from the growing concern, within the music industry and among listeners, regarding the integrity of vocal performances. In an era where technology can significantly alter artistic expression, understanding and detecting the use of tools like Auto-Tune becomes crucial for preserving the authenticity and credibility of musical compositions.
In this study, a data-driven approach for detecting Auto-Tuned vocals in music is presented, backed by our creation of a unique dataset with original and Auto-Tuned audio. Our proposed methodology utilizes triplet networks to emphasize distinctions between Auto-Tuned and genuine vocal performances, leveraging mel-spectrograms as the primary input to the model. This decision is backed by the mel-spectrogram's capacity to closely align with human auditory perception, capturing crucial frequency information while addressing the nuances of vocal characteristics essential for distinguishing between Auto-Tuned and authentic vocals. To the best of our knowledge, no prior effort has been specifically made in this direction. Nevertheless, we adopted a variant of Rawnet2 \cite{rawnet2}, designed for audio forgery detection, as a baseline. This network has been proposed by Tak et al. \cite{Tak2021} to enhance Rawnet2 effectiveness in anti-spoofing and audio forgery detection. Due to its strong performance, this method was also used as a baseline in the ASVspoof 2021 challenge \citep{asvspoof2021}, a prominent competition focused on spoofed and deepfake speech detection. The implementation code utilized in this research is available on GitHub\footnote{\url{https://mahyargm.github.io/Auto-Tune-Detection}}.

The remainder of the paper is organized as follows. Section \ref{sec:method} outlines the problem formulation and details the Auto-Tune detection pipeline. Section \ref{sec:dataset}, accounts for the dataset creation process. Section \ref{sec:experiments} describes the experimental settings and evaluates the performance and validity of our approach through experiments. Finally, we draw our conclusions in Section \ref{sec:discussion}.

\section{Auto-Tune detection}
\label{sec:method}

This section explores Auto-Tune detection by first formulating the problem and its challenges. Then, the image processing approach, used to extract features from mel-spectrogram representation of audio clips and classify them, is presented.

\subsection{Problem Formulation}

Auto-tuning consists of editing a music piece to alter or correct the pitch of a vocalist's singing. When a singer hits a note slightly off-key, Auto-Tune can automatically adjust the pitch to bring it into tune. This can help smooth out imperfections in a vocal performance and create a more polished final product.
The Auto-Tune detection process identifies segments or patterns in the audio signal where the pitch shows unnatural consistency or manipulation, deviating significantly from the expected fluctuations in a genuine performance. Specifically, our focus lies in identifying instances where auto-tuning has been extensively applied to alter the original vocals, impacting the authenticity and natural voice of the singer.
 
Formally, let us define a digital audio signal representing a song as $\x = \v + \a$, where $\v$ is the \emph{vocal} component (i.e., the main singing voice) and $\a$ is the \emph{accompaniment} (i.e., remaining musical background).
Our goal is to develop a model $\M$ that returns $\{y_i\}_{i=1}^n = \M(\x)$, where $n$ is the number of segments in the song and $y_i \in [0, 1]$ is the likelihood of the $i$th segment of the song $\x$ being Auto-Tuned.

\subsection{Methodology}

\begin{figure}[t]
  \centerline
  {\includegraphics[width=0.5\columnwidth]{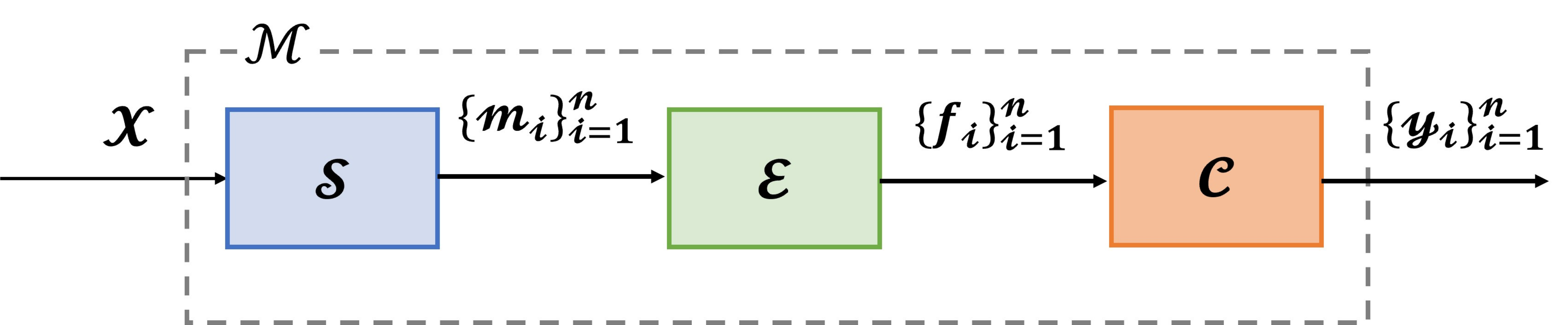}}
  \caption{Pipeline of the proposed method.}
  \label{fig:pipeline}
\end{figure}

Auto-tuning is usually applied in post-production, making it difficult for listeners to detect its presence in a song without access to the raw stems. In addition, the parameters of auto-tuning tools, such as correction speed and scale, can vary widely among different songs and genres. These variations make it challenging to define a one-size-fits-all approach to detect Auto-Tuned vocals. As shown in Fig. \ref{fig:pipeline}, the proposed data-driven method is composed of three main blocks: an audio pre-processing block $\S$; a feature extraction block $\E$; and a classification block $C$. In the following, we report the details of each step.

\vspace{.5em}\textbf{Pre-Processing.}
Since auto-tuning is applied to a single vocal track rather than a full song, verifying the authenticity of a performance requires examining the isolated vocal. Therefore, a separation tool is necessary to distinguish the singing from the accompaniment. In this study, vocal isolation was accomplished using an available implementation of Vocal-Remover \cite{vocalremover}. For any song $\x$, the vocal isolation tool estimates the vocal component $\v$.

The next step involves segmenting the vocal part $\v$ into $n$ fixed-size segments $\v_i$. To do so, we divide the input into fixed-size segments (10 seconds in our experiments) and then filter out segments with less energy than 0.002\% of the maximum energy. In this way, we discard segments devoid of vocal elements.

\begin{figure}[t]
  \centerline
  {\includegraphics[width=\columnwidth]{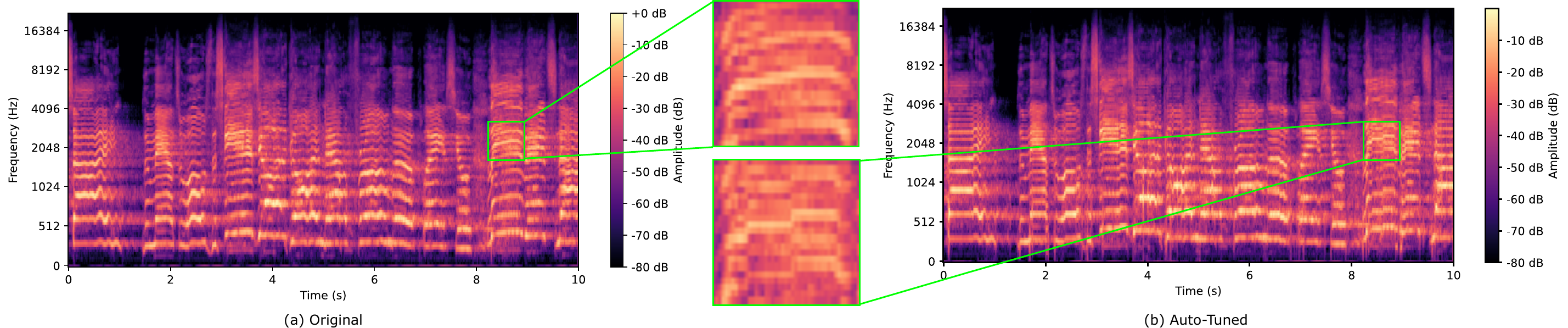}}
  \caption{The spectrogram of a 10-second vocal (a) and its corresponding Auto-Tuned version (b).}
  \label{fig:spectrograms}
\end{figure}

The mel-spectrogram is known to be suitable for highlighting subtle information in the audio signal (see Fig.~\ref{fig:spectrograms}). Consequently, the next stage involves converting the audio segment under analysis $\v_i$ into its mel-spectrogram representation $\m_i$. To do so, considering a sampling rate of 44.1 kHz, the spectrogram is computed using frames of size 2048 samples extracted with a hop size of 1024 samples. By employing 128 mel-frequency bins, this configuration offers a finely detailed representation of the spectral characteristics of the audio signal.

\vspace{.5em}\textbf{Feature Extraction.}

The feature extractor $\E$ is composed of a standard convolutional neural network (CNN) architecture (backbone) followed by a fully connected layer to generate feature vectors $\f_i=\E(\m_i)$ of size 512 for $i \in \{1,2,...,n\}$. In our approach, a triplet network \cite{Triplet} has been employed to train the feature extractor. Input triplets are generated through the Semi-Hard triplet selection method \cite{semi-hard} to enhance the convergence of the model.

\vspace{.5em}\textbf{Classification.}
The extracted feature vectors $\{\f_i\}_{i=1}^n$ are then used as input to a two-layer fully connected binary classifier $\C$ to predict if each segment of the audio track is Auto-Tuned or not. Formally we compute $y_i=\C(\f_i)$ for $i \in \{1,2,...,n\}$, which represents the likelihood of segment $\x_i$ being Auto-Tuned. The track under analysis $\x$ is classified as Auto-Tuned only if at least a certain number of segments are predicted as Auto-Tuned. The threshold selection is contingent on the specific application requirements and the desired level of confidence.

\section{Dataset}
\label{sec:dataset}

\begin{figure}[t]
  \centering
  \includegraphics[width=.5\columnwidth]{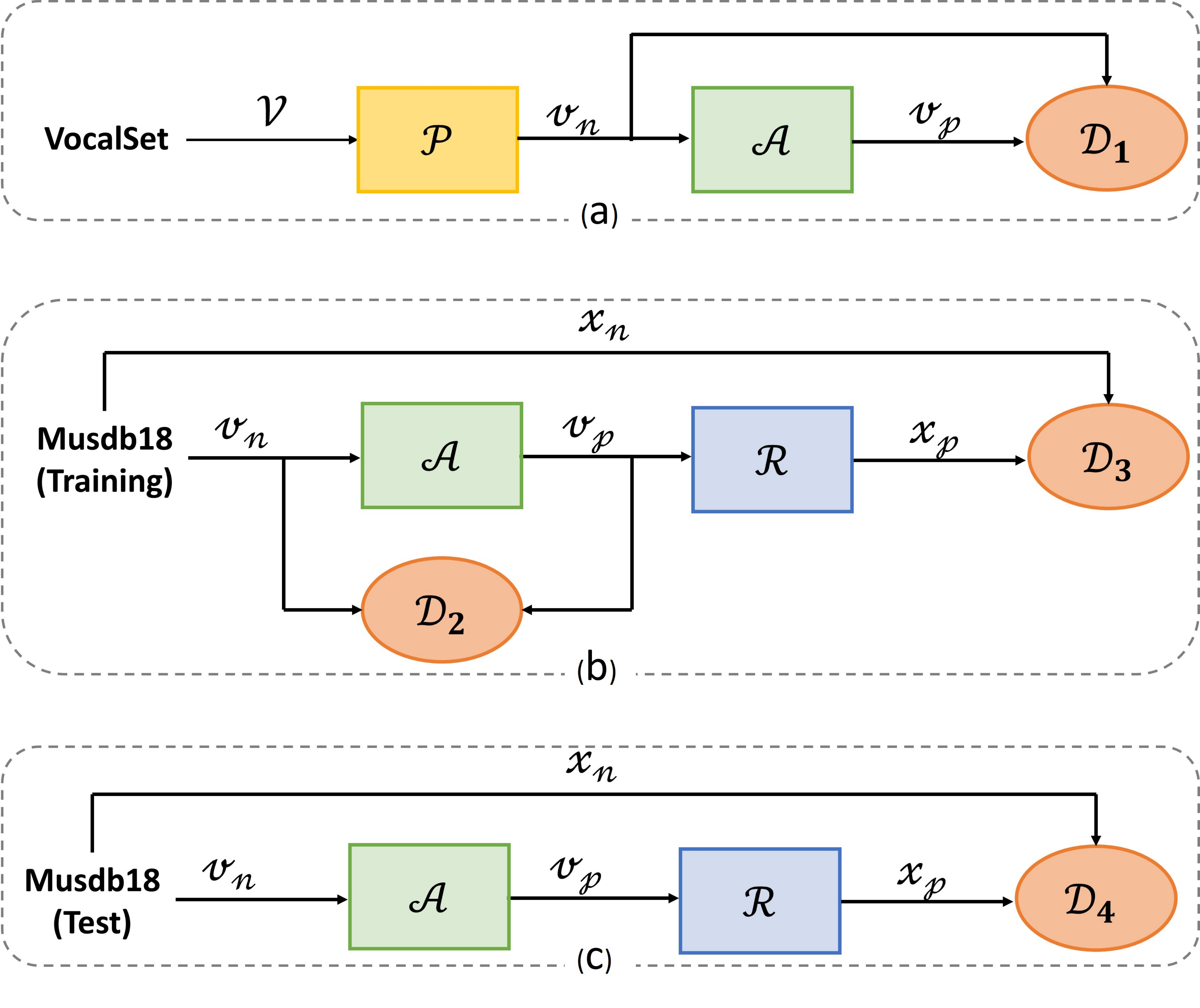}
  \caption{The schematic of dataset creation pipelines: $\D_1$ derived from VocalSet (a), $\D_2$ and $\D_3$ derived from Musdb18 (b), and the test dataset ($\D_4$) creation process (c).}
  \label{fig:dataset}
\end{figure}

Since there are no pre-existing datasets containing both Auto-Tuned musical compositions and authentic ones, we have decided to generate one for this and potential future research in the field. Owing to copyright considerations, we decided to use two existing datasets in MIR as a starting point instead of creating them from scratch. VocalSet \cite{vocalset} and Musdb18 \cite{musdb18} have been employed for this purpose.
In the remainder of this section, we first introduce the used processing blocks and then we detail the dataset creation process summarized in Fig. \ref{fig:dataset}. 

\subsection{Pitch correction process}
\label{pitchcorr}

Pitch correction is applied to the vocal part $\v$ of a song to generate the Auto-Tuned signal $\vp = \A(\v)$. The first step in performing pitch correction involves determining the current pitch at a specific time point, a problem commonly referred to as pitch tracking in Digital Signal Processing (DSP). Among available methods, a notable and widely recognized approach is the PYIN algorithm \cite{PYIN}. The foundation of PYIN lies in the YIN algorithm \cite{YIN}. The YIN algorithm focuses on pitch estimation by analyzing the autocorrelation function of the time-domain signal and subsequently refining the obtained results. PYIN advances this method by incorporating a Hidden Markov Model into the outcomes produced by the YIN algorithm.

The pitch of each frame is estimated using the PYIN algorithm, and subsequently is transposed to the nearest MIDI note by applying the pitch-synchronized overlap and add technique (PSOLA) \cite{PSOLA}. While this process may not replicate the precise procedures employed in professional auto-tuning of musical compositions, its purpose is to generate comparable artifacts, making it a valuable tool for this research.

\subsection{Dataset creation process}

The training dataset comprises three distinct subsets, denoted as $\D_1$, $\D_2$, and $\D_3$, while the test dataset is $\D_4$.

The first dataset, denoted as $\D_1$, is produced starting from VocalSet which contains 10.1 hours of monophonic recorded audio of professional singers demonstrating both standard and extended vocal techniques. These recordings are from twenty different performers (nine male, eleven female). Due to varied lengths ranging from 1.56 to 81.21 seconds (with an average duration of 8.76 seconds), we standardized the audio by either trimming or padding ($\P$) to create 10-second audio files. Given that VocalSet exclusively comprises vocal elements without additional instrumentation, pitch correction is applied directly to each original audio recording $\mathbf{v}_\text{n}$ (i.e., negative class) to create auto-tuned samples $\mathbf{v}_\text{p}$ (i.e., positive class). The $\D_1$ dataset is comprised of the resultant paired instances.

The Musdb18 dataset, the other primary dataset used in this research, is comprised of 100 professionally recorded songs allocated for training and 50 for testing purposes. This dataset includes songs alongside their corresponding isolated vocal tracks. Each isolated vocal $\mathbf{v}_\text{n}$ existing in the training part of the Musdb18 dataset undergoes auto-tuning to generate its corresponding Auto-Tuned sample $\mathbf{v}_\text{p} = \A(\mathbf{v}_\text{n})$. These audio pairs form the dataset $\D_2$. Additionally, to simulate a more realistic scenario for Auto-Tune detection, positive samples $\mathbf{v}_\text{p}$ are recombined with their associated accompaniment elements $\mathbf{a}$, which are available in the Musdb18 dataset, to create Auto-Tuned songs $\mathbf{x}_\text{p} = \R(\mathbf{v}_\text{p}, \mathbf{a})$. These Auto-Tuned songs along with their corresponding authentic songs $\mathbf{x}_\text{n}$ constitute the dataset $\D_3$.
The methodology employed to build $\D_3$ is used also for the test dataset $\D_4$, utilizing the test partition of Musdb18. This approach aimed to provide a more authentic evaluation of the model in real-world cases, emphasizing scenarios where the raw stems of the song are not accessible. 

The ultimate training dataset comprises 3613 pairs of 10-second vocal performances obtained from VocalSet, alongside 100 pairs of composed songs and 100 pairs of isolated vocals. The test dataset consists of 50 pairs of composed songs.

\begin{figure*}[t]
  \centering
  \includegraphics[width=\textwidth]{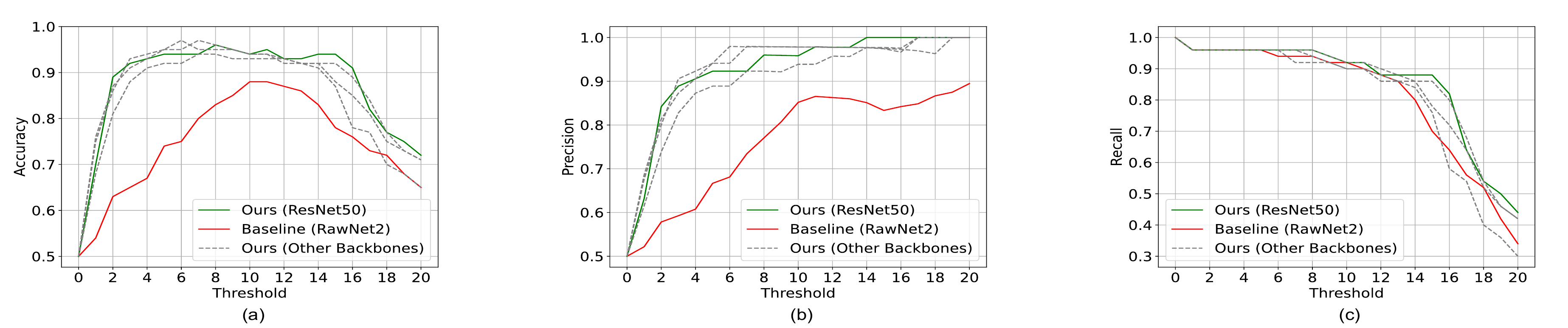}
  \caption{Song-level performance curves of the models across varying thresholds, measured in terms of accuracy (a), precision (b), and recall (c).}
  \label{fig:threshold}
\end{figure*}

\section{Experiments}
\label{sec:experiments}

In this section, we assess the accuracy and efficiency of our approach. First, we outline the experimental configurations selected for training and evaluating our model. Then, we provide a comprehensive analysis of the experimental results.

\subsection{Experimental settings}

As shown in Section \ref{sec:method}, our detection method is built on top of two networks: the triplet network $\E$ and the classifier $\C$.

As a backbone for the triplet network, we adopted well-known architectures: ResNet18, ResNet50 \cite{ResNet}, ResNeXt-50 (32×4d) \cite{ResNeXt}, and EfficientNetv2 (small) \cite{EfficientNet}. All backbones are pre-trained on ImageNet \cite{imagenet}. This approach showcases the framework's independence from specific model architectures. 

The triplet network processes sets of mel-spectrograms sized at 431 by 128, generating feature vectors of size 512 for each spectrogram. Subsequently, the model optimization was executed using the triplet loss function. During the triplet network training phase, a batch size of 64 and a learning rate of $10^{-4}$ were employed.    
The classifier is composed of two fully connected layers, with dimensions 256 and 64 respectively. These layers utilize the ReLU activation function. Additionally, the output layer employs a Sigmoid activation function to produce the final output. The model underwent training until convergence on a validation set, employing a batch size of 64 and a learning rate set at $10^{-5}$, and the optimization utilized binary cross-entropy loss during the training process. 

Given the absence of prior work addressing Auto-Tune detection, we opted to utilize Tak et al. \cite{Tak2021} work, which is based on RawNet2 \cite{rawnet2} and originally designed for anti-spoofing, as our baseline for this task. The architecture of RawNet2 involves CNNs and self-attention mechanisms. CNNs are particularly effective in capturing local patterns and features, while self-attention mechanisms enable the model to weigh different parts of the input sequence differently, focusing on relevant segments for better feature extraction. 
\subsection{Experimental Results}
In this work, we trained this architecture on our dataset to process raw 10-second audio segments and predict whether they are Auto-Tuned or not. Henceforth, in the remainder of the paper, this model will be referred to as the Baseline.
The Baseline has undergone training for a maximum of 100 epochs and with a learning rate set at $10^{-4}$ and a batch size of 16, during which the best model, characterized by the lowest validation loss, has been selected. The model was optimized using binary cross-entropy loss.

\begin{table}
    \caption{Segment-level results on the test dataset.}
    \label{tab:segment-level}
    \centering
    \footnotesize
    \begin{tabular}{|c|c|c|c|c|}
    \hline
    \textbf{Approach} & \textbf{Backbone} & \textbf{Precision} & \textbf{Recall} & \textbf{Accuracy} \\
    \hline
    Baseline & RawNet2 & 74.03\% & 90.35\% & 79.33\%\\
    Ours & ResNet18 & \textbf{93.56\%} & 88.46\% & 91.19\%\\
    Ours & ResNet50 & 93.04\% & \textbf{96.75\%} & \textbf{94.75\%}\\
    Ours & ResNeXt & 90.09\% & 96.33\% & 92.86\%\\
    Ours & EfficientNet & 93.38\% & 93.18\% & 93.28\%\\
    \hline
    \end{tabular}
\end{table}
\subsubsection{Segment-level results}
To assess the effectiveness of our method on 10-second segments, we tested our approach utilizing ResNet50, ResNet18, ResNeXt, and EfficientNetv2 (small) as backbones.

We then compared these results with the results obtained by applying the trained Baseline.

The findings presented in Table. \ref{tab:segment-level} illustrate the superiority of our approach over the Baseline, particularly in terms of precision and accuracy.

\subsubsection{Song-level results}

In practical scenarios, the objective of Auto-Tune detection extends to determining whether an entire song has been subjected to auto-tuning. Hence, our approach was assessed at the song-level. To achieve this, all segments of each song underwent processing by the networks, and the count of segments identified as Auto-Tuned was calculated. Subsequently, if this count surpassed a certain threshold, we labeled the entire song as Auto-Tuned.

The performance of both the baseline and our models across various thresholds is depicted in Fig. \ref{fig:threshold}.

\subsubsection{Robustness evaluation}

\begin{table}
    \caption{Accuracy of models in the presence of post-processing techniques and MP3 compression.}
    \footnotesize
    \setlength{\tabcolsep}{3pt}
    \label{tab:robustness}
    \centering
    \begin{tabular}{|c|c|c|c|c|}
    \hline
    \textbf{Approach} & \textbf{Backbone} & \textbf{MP3 Compression} & \textbf{Random processing} \\
    \hline
    Baseline & RawNet2 & 56.67\% & 74.71\%\\
    Ours & ResNet18 & 82.46\% & 86.61\%\\
    Ours & ResNet50 & 80.99\% & 89.22\%\\
    Ours & ResNeXt & \textbf{87.55\%} & 84.90\%\\
    Ours & EfficientNet & 87.34\% & \textbf{90.72\%}\\
    \hline
    \end{tabular}
\end{table}

In many cases, Auto-Tuned songs undergo post-processing that can render them challenging to recognize. To evaluate the robustness of the trained models against these post-processing alterations, two additional experiments were conducted across all models.

First, we evaluated the performance of the models on the test dataset after the application of lossy MP3 compression with the rate randomly selected between 32 and 64 kbps.

Second, the following modifications were applied to all the songs in the test dataset each of them with a probability of 50\%:
\begin{itemize}[leftmargin=*]
    \item Addition of Gaussian noise with the amplitude randomly chosen between 0.001 and 0.015.
    \item Random variations in speed, allowing for a maximum increase of 25\% or decrease of 20\% in the audio file's speed.
    \item Random shifting of the sound either forward or backward within a range of -0.5 to 0.5 seconds.
\end{itemize}

Table. \ref{tab:robustness} compares and illustrates the accuracy of trained models after applying the above-mentioned post-processing techniques to the test dataset.

\subsubsection{Real-world use-case}
As a last experiment, we run the proposed detector on a few songs that are affected by different degrees of auto-tune. Results on Github\footnote{\url{https://mahyargm.github.io/Auto-Tune-Detection}} show that the detector performs well also in the wild.

\section{Discussion}
\label{sec:discussion}

Our study represents a pioneering effort in the domain of Auto-Tune detection, as no prior work has specifically addressed this challenge. To enhance the strength of our investigation, we created a new dataset dedicated to Auto-Tune detection, leveraging existing datasets in Music Information Retrieval (MIR).

In the absence of established benchmarks for Auto-Tune detection, we chose to evaluate our approach against the Rawnet2 architecture, a well-known model in audio processing and audio forgery detection. Our results demonstrate the superiority of the proposed method, surpassing Rawnet2 in both accuracy and robustness. This achievement underscores the effectiveness of our data-driven approach and highlights its potential for advancing the field of music forgery detection.

While our study marks a significant step, it is essential to acknowledge certain limitations. Notably, the absence of manually and professionally Auto-Tuned songs in our dataset represents a constraint. Although we successfully replicated auto-tuning behavior using the employed method, the inclusion of manually and professionally Auto-Tuned samples would have provided a more comprehensive evaluation of our model's performance. Future iterations of our research should aim to incorporate such samples.

{\small
\bibliographystyle{plainnat}
\bibliography{ref}

\begin{thebibliography}{25}
\providecommand{\natexlab}[1]{#1}
\providecommand{\url}[1]{\texttt{#1}}
\expandafter\ifx\csname urlstyle\endcsname\relax
  \providecommand{\doi}[1]{doi: #1}\else
  \providecommand{\doi}{doi: \begingroup \urlstyle{rm}\Url}\fi

\bibitem[{Antares Audio Technologies}(2019)]{antares-autotune}
{Antares Audio Technologies}.
\newblock {Antares Auto-Tune 9}.
\newblock Software, 2019.
\newblock URL \url{{https://www.antarestech.com/product/auto-tune/}}.

\bibitem[Bhagtani et~al.(2022)Bhagtani, Yadav, Bartusiak, Xiang, Shao, Baireddy, and Delp]{bhagtani2022}
K.~Bhagtani, A.K.S. Yadav, E.R. Bartusiak, Z.~Xiang, R.~Shao, S.~Baireddy, and E.J. Delp.
\newblock An overview of recent work in multimedia forensics.
\newblock In \emph{2022 IEEE 5th International Conference on Multimedia Information Processing and Retrieval (MIPR)}, pages 324--329. IEEE Computer Society, August 2022.

\bibitem[Borkar et~al.(2021)Borkar, Patre, Khalsa, Kawale, and Chakurkar]{borkar2021}
N.~Borkar, S.~Patre, R.S. Khalsa, R.~Kawale, and P.~Chakurkar.
\newblock Music plagiarism detection using audio fingerprinting and segment matching.
\newblock In \emph{2021 Smart Technologies, Communication and Robotics (STCR)}, pages 1--4. IEEE, October 2021.

\bibitem[Cheveigné and Kawahara(2002)]{YIN}
Alain~De Cheveigné and Hideki Kawahara.
\newblock Yin, a fundamental frequency estimator for speech and music.
\newblock \emph{The Journal of the Acoustical Society of America}, 111\penalty0 (4):\penalty0 1917--1930, 2002.

\bibitem[Deng et~al.(2009)Deng, Dong, Socher, Li, Li, and Fei-Fei]{imagenet}
J.~Deng, W.~Dong, R.~Socher, L.J. Li, K.~Li, and L.~Fei-Fei.
\newblock Imagenet: A large-scale hierarchical image database.
\newblock In \emph{2009 IEEE Conference on Computer Vision and Pattern Recognition}, pages 248--255. IEEE, June 2009.

\bibitem[Gurjar et~al.(2023)Gurjar, Moon, and Abuhmed]{gurjar2023}
K.~Gurjar, Y.S. Moon, and T.~Abuhmed.
\newblock Trumuzic: A deep learning and data provenance-based approach to evaluating the authenticity of music.
\newblock \emph{Applied Sciences}, 13\penalty0 (16):\penalty0 9425, 2023.

\bibitem[He et~al.(2016)He, Zhang, Ren, and Sun]{ResNet}
Kaiming He, Xiangyu Zhang, Shaoqing Ren, and Jian Sun.
\newblock Deep residual learning for image recognition.
\newblock In \emph{Proceedings of the IEEE Conference on Computer Vision and Pattern Recognition}, 2016.

\bibitem[He et~al.(2021)He, Liu, Gong, Yan, and Zhang]{he2021}
T.~He, W.~Liu, C.~Gong, J.~Yan, and N.~Zhang.
\newblock Music plagiarism detection via bipartite graph matching.
\newblock \emph{arXiv preprint arXiv:2107.09889}, 2021.

\bibitem[Hoffer and Ailon(2015)]{Triplet}
E.~Hoffer and N.~Ailon.
\newblock Deep metric learning using triplet network.
\newblock In \emph{Similarity-Based Pattern Recognition: Third International Workshop, SIMBAD 2015, Copenhagen, Denmark, October 12-14, 2015. Proceedings}, volume~3. Springer International Publishing, 2015.

\bibitem[Jung et~al.(2020)Jung, Kim, Shim, Kim, and Yu]{rawnet2}
Jee-weon Jung, Seung-bin Kim, Hye-jin Shim, Ju-ho Kim, and Ha-Jin Yu.
\newblock Improved rawnet with feature map scaling for text-independent speaker verification using raw waveforms.
\newblock \emph{Proc. Interspeech}, pages 3583--3587, 2020.

\bibitem[Li et~al.(2013)Li, Zhang, and Wang]{wli2013}
W.~Li, X.~Zhang, and Z.~Wang.
\newblock Music content authentication based on beat segmentation and fuzzy classification.
\newblock \emph{EURASIP Journal on Audio, Speech, and Music Processing}, pages 1--13, 2013.

\bibitem[Liu et~al.(2023)Liu, Wang, Sahidullah, Patino, Delgado, Kinnunen, Todisco, Yamagishi, Evans, Nautsch, and Lee]{asvspoof2021}
X.~Liu, X.~Wang, M.~Sahidullah, J.~Patino, H.~Delgado, T.~Kinnunen, M.~Todisco, J.~Yamagishi, N.~Evans, A.~Nautsch, and K.A. Lee.
\newblock Asvspoof 2021: Towards spoofed and deepfake speech detection in the wild.
\newblock \emph{IEEE/ACM Transactions on Audio, Speech, and Language Processing}, 2023.

\bibitem[López-García et~al.(2022)López-García, Martínez-Rodríguez, and Liern]{lopezgarcia2022}
A.~López-García, B.~Martínez-Rodríguez, and V.~Liern.
\newblock A proposal to compare the similarity between musical products. one more step for automated plagiarism detection?
\newblock In \emph{International Conference on Mathematics and Computation in Music}, pages 192--204, Cham, June 2022. Springer International Publishing.

\bibitem[Malandrino et~al.(2022)Malandrino, Prisco, Ianulardo, and Zaccagnino]{malandrino2022}
D.~Malandrino, R.~De Prisco, M.~Ianulardo, and R.~Zaccagnino.
\newblock An adaptive meta-heuristic for music plagiarism detection based on text similarity and clustering.
\newblock \emph{Data Mining and Knowledge Discovery}, 36\penalty0 (4):\penalty0 1301--1334, 2022.

\bibitem[Mauch and Dixon(2014)]{PYIN}
Matthias Mauch and Simon Dixon.
\newblock pyin: A fundamental frequency estimator using probabilistic threshold distributions.
\newblock In \emph{2014 IEEE International Conference on Acoustics, Speech and Signal Processing (ICASSP)}. IEEE, 2014.

\bibitem[Park et~al.(2022)Park, Baek, Jeon, and Jeong]{park2022}
K.~Park, S.~Baek, J.~Jeon, and Y.S. Jeong.
\newblock Music plagiarism detection based on siamese cnn.
\newblock \emph{Hum.-Cent. Comput. Inf. Sci}, 12:\penalty0 12--38, 2022.

\bibitem[Prisco et~al.(2017)Prisco, Malandrino, Zaccagnino, and Zaccagnino]{deprisco2017}
R.~De Prisco, D.~Malandrino, G.~Zaccagnino, and R.~Zaccagnino.
\newblock Fuzzy vectorial-based similarity detection of music plagiarism.
\newblock In \emph{2017 IEEE International Conference on Fuzzy Systems (FUZZ-IEEE)}, pages 1--6. IEEE, July 2017.

\bibitem[Rafii et~al.(2017)Rafii, Liutkus, St{\"o}ter, Mimilakis, and Bittner]{musdb18}
Zafar Rafii, Antoine Liutkus, Fabian-Robert St{\"o}ter, Stylianos~Ioannis Mimilakis, and Rachel Bittner.
\newblock The {MUSDB18} corpus for music separation, December 2017.
\newblock URL \url{https://doi.org/10.5281/zenodo.1117372}.

\bibitem[Schroff et~al.(2015)Schroff, Kalenichenko, and Philbin]{semi-hard}
Florian Schroff, Dmitry Kalenichenko, and James Philbin.
\newblock Facenet: A unified embedding for face recognition and clustering.
\newblock In \emph{Proceedings of the IEEE Conference on Computer Vision and Pattern Recognition (CVPR)}, 2015.

\bibitem[Tak et~al.(2021)Tak, Patino, Todisco, Nautsch, Evans, and Larcher]{Tak2021}
H.~Tak, J.~Patino, M.~Todisco, A.~Nautsch, N.~Evans, and A.~Larcher.
\newblock End-to-end anti-spoofing with rawnet2.
\newblock In \emph{ICASSP 2021-2021 IEEE International Conference on Acoustics, Speech and Signal Processing (ICASSP)}, pages 6369--6373. IEEE, June 2021.

\bibitem[Tan and Le(2019)]{EfficientNet}
M.~Tan and Q.~Le.
\newblock Efficientnet: Rethinking model scaling for convolutional neural networks.
\newblock In \emph{International Conference on Machine Learning}, pages 6105--6114. PMLR, May 2019.

\bibitem[{tsurumeso}()]{vocalremover}
{tsurumeso}.
\newblock {vocal-remover}.
\newblock \url{https://github.com/tsurumeso/vocal-remover}.
\newblock Accessed on 2023-12-10.

\bibitem[Valbret et~al.(1992)Valbret, Moulines, and Tubach]{PSOLA}
Hélene Valbret, Eric Moulines, and Jean-Pierre Tubach.
\newblock Voice transformation using psola technique.
\newblock \emph{Speech Communication}, 11\penalty0 (2-3):\penalty0 175--187, 1992.

\bibitem[Wilkins et~al.(2018)Wilkins, Seetharaman, Wahland, and Pardo]{vocalset}
J.~Wilkins, Prem Seetharaman, Alison Wahland, and Bryan Pardo.
\newblock Vocalset: A singing voice dataset, 2018.
\newblock URL \url{https://doi.org/10.5281/zenodo.1442513}.

\bibitem[Xie et~al.(2017)Xie, Girshick, Dollár, Tu, and He]{ResNeXt}
S.~Xie, R.~Girshick, P.~Dollár, Z.~Tu, and K.~He.
\newblock Aggregated residual transformations for deep neural networks.
\newblock In \emph{Proceedings of the IEEE Conference on Computer Vision and Pattern Recognition}, pages 1492--1500, 2017.

\end{thebibliography}
}
\end{document}